\documentclass[twocolumn,prb,showpacs,aps,,superscriptaddress]{revtex4-1}

\usepackage[english]{babel}
\usepackage{times}
\usepackage{graphicx}
\usepackage{graphics}
\usepackage{amsmath}
\usepackage{amsfonts}
\usepackage{amssymb}
\usepackage{epstopdf}
\usepackage{makeidx}
\usepackage{subfigure}
\usepackage{color}
\usepackage{pgf}
\usepackage{bm}
\usepackage{tikz} % Create PostScript and PDF graphics in TeX. Depends on the PGF package.
\usepackage[normalem]{ulem}
\usepackage{soul}
\newcommand{\be}{\begin{equation}}
\newcommand{\ee}{\end{equation}}
\newcommand{\bea}{\begin{eqnarray}}
\newcommand{\eea}{\end{eqnarray}}

\makeindex
\begin{document}
%\doublespacing
%\title{Quantum Hall Effect in Graphene with Interface-Induced Spin-Orbit Coupling}
%

\title{Tuning quantum reflection in graphene with an external magnetic field}
\author{Marcius Silvestre}
\affiliation{Instituto de F\'{\i}sica, Universidade Federal do Rio de Janeiro,
Caixa Postal 68528, Rio de Janeiro 21941-972, RJ, Brazil}

\author{Tarik P. Cysne}
\email{tarik.cysne@gmail.com}
\affiliation{Instituto de F\'{\i}sica, Universidade Federal do Rio de Janeiro,
Caixa Postal 68528, Rio de Janeiro 21941-972, RJ, Brazil}

\author{Daniela Szilard}
\affiliation{Instituto de F\'{\i}sica, Universidade Federal do Rio de Janeiro,
Caixa Postal 68528, Rio de Janeiro 21941-972, RJ, Brazil}
\affiliation{Centro Brasileiro de Pesquisas Físicas, Rua Dr. Xavier Sigaud, 150 - Urca - Rio de Janeiro - RJ - Brasil - CEP: 22290-180}
\author{Felipe A. Pinheiro}
\affiliation{Instituto de F\'{\i}sica, Universidade Federal do Rio de Janeiro,
Caixa Postal 68528, Rio de Janeiro 21941-972, RJ, Brazil}
\author{Carlos Farina}
\affiliation{Instituto de F\'{\i}sica, Universidade Federal do Rio de Janeiro,
Caixa Postal 68528, Rio de Janeiro 21941-972, RJ, Brazil}
\date{\today}

\begin{abstract}

We theoretically demonstrate that an external magnetic field can be used to control quantum reflection of matter waves in graphene due to its extraordinary magneto-optical properties. We calculate the quantum reflection probabilities in graphene for three experimentally relevant atomic species (He, Na, and Rb) using the full Casimir-Polder potential computed by Lifshitz formula valid at all distance regimes, going beyond the traditional approach to quantum reflection, based on power law potentials, which are known to be valid only in the short distance (non-retarded van der Waals) or in the large distance (retarded) regimes. We predict the energy range for which quantum reflection is more likely to occur as a function of the magnetic field, and show that the quantum reflection probabilities exhibit discontinuities that reflect the structure of Landau levels in graphene. Altogether our findings suggest an alternative way to control quantum reflection at the nanoscale, and pave the way for the design of alternative, magnetically tuned reflective diffraction elements for matter waves.

 \end{abstract}

\maketitle
%	
%{\footnotesize *These authors contributed equally to this work and are joint first authors.}
% ---------------------------------------------------------
%
\section{Introduction}

One of the most curious and unexpected effects related to the wave nature of quantum particles is the so called Quantum Reflection (QR). It consists in the reflection of a quantum particle that moves under the action of a potential that decreases monotonically in the direction of the particle motion, even without the existence of any turning point. Classically, the particle would suffer a force pushing it in the forward direction. Rather than, increasing its velocity, the particle is reflected by such a decreasing potential. The probability of occurrence of QR increases as the wave nature of the particles becomes more pronounced\cite{DeBroglie}, facilitating QR of particles with low masses at low energies. As a result there is a non-zero probability of an atom at low energies, attracted by a wall due to dispersive forces, to be reflected before reaching the wall. It is worth mentioning that this intriguing phenomenon is not restricted to quantum mechanics.  Indeed, it is a general feature of wave propagation in inhomogeneous media and it may occur for mechanical waves or electromagnetic waves in dielectrics and transmission lines\cite{Brekhovskikh-2012}. 

The first experiments on QR were performed with helium and hydrogen atoms reflected by liquid helium surfaces \cite{Nayak-1983, Berkhout-1989, Doyle-1991, Yu-1993}. Due to the low mass of  atomic specimens in these experiments, QR regime was reached with relatively high energies ($1-10 neV$). Heavier the particle, lower its energy should be in order to reach the QR regime. In references \cite{Pasquini-2004, Pasquini-2006, Marchant-2016} deep QR regimes have been reached for sodium and rubidium atoms prepared in Bose-Einstein states with normal incidence and energies of the order of $10^{-4}neV$. Lower energy regimes can be reached with oblique incidence of the incoming particles \cite{Druzhinina-2003, Zhao-2008}. Since the pioneering paper by Shimizu \cite{Shimizu}, in which an ingenious setup based on QR was developed to investigate the power laws of the non-retarded and retarded dispersive interactions between an atom and a wall, many different approaches have been put forward to probe dispersive forces via QR\cite{Oberst-2005, Zhao-2008, Zhao-2010, Zhao-2013, Barnea-2017}. In this context, developing alternative mechanisms to control and/or tuning the probability  of a beam of atoms to be reflected by a wall may open new possibilities for designing new atomic mirrors \cite{Cote-2003, Segev-1997} or even  atomic traps \cite{Hammes-2002, Crepin-2017, Jurisch-2008}.  Quantum reflection plays an important role in many other areas of physics, which include atom optics \cite{Cronin-RMP-2009, Deutschmann-1993, Landragin-1997, Shimizu-Fujita-2002, Kohno-2003, Savalli-2002}  and, more recently, in high precision measurements of the short-range regime of gravitational forces \cite{Perez-2015, Dufour-2014}.

Quantum reflection strongly depends on the Casimir-Polder interaction between the incident particles and the reflecting wall. This interaction may change substantially if one changes the material properties of the wall and the type of incident particles. In this context 2D materials, such as graphene, emerged as good candidates control QR due to their remarkable electromagnetic and mechanical properties. Indeed, it has been shown that interactions mediated by vacuum fluctuations in graphene are highly tunable by varying the chemical-potential \cite{Cysne-2016, Bordag-2016, Henkel-2018}, external magnetic field \cite{Cysne-Kort-Kamp_2014, Macdonald_PRL}, strain\cite{Nichols-2016}, and by stacking of many graphene sheets \cite{Khusnutdinov-2016, Khusnutdinov-2018}. Regarding QR by graphene sheets, some theoretical works do exist~\cite{Judd-2011}, but only very few of them explore the important application of tuning this effect using external parameters, such as strain~\cite{Nichols-2016}. Taking advantage of the remarkable magneto-optical control of the Casimir-Polder interaction between atoms and graphene sheets \cite{Cysne-Kort-Kamp_2014}, we put forward an alternative, realistic method for tuning QR of cold atoms by a graphene sheet by applying a perpendicular magnetic field. We demonstrate that the quantum reflection probability for a given energy as a function of the external magnetic field shows discontinuities, a direct consequence of the structure of the Landau levels of the electronic spectrum of the graphene sheet.
 
This paper is organized as follows: in the next section we present the methodology to be used in the computation of QR probabilities. However, instead of using power law expressions for the dispersive forces which are valid only in the two opposite regimes of short distance (van der Waals regime) and large distance regime  (retarded regime) we use the complete Casimir-Polder potential computed by means of Lifshitz formula \cite{Lifshitz-1956, Milonni}, valid for all distance regimes. In Section III we present our results and show that graphene is, indeed, a very good platform to be used as reflecting  material and permits high tunability of QR. We discuss QR for three atoms with different masses (He, Na, Rb) and show that depending on the atom mass its QR probability will be significant for different atom-graphene distance regimes. Section IV is left for conclusions and final remarks. 

%-----------------------------------------------------------
\section{Methodology}

\begin{figure}[h]
%\vspace{0.1in}
  \centering
  \includegraphics[scale = 0.7]{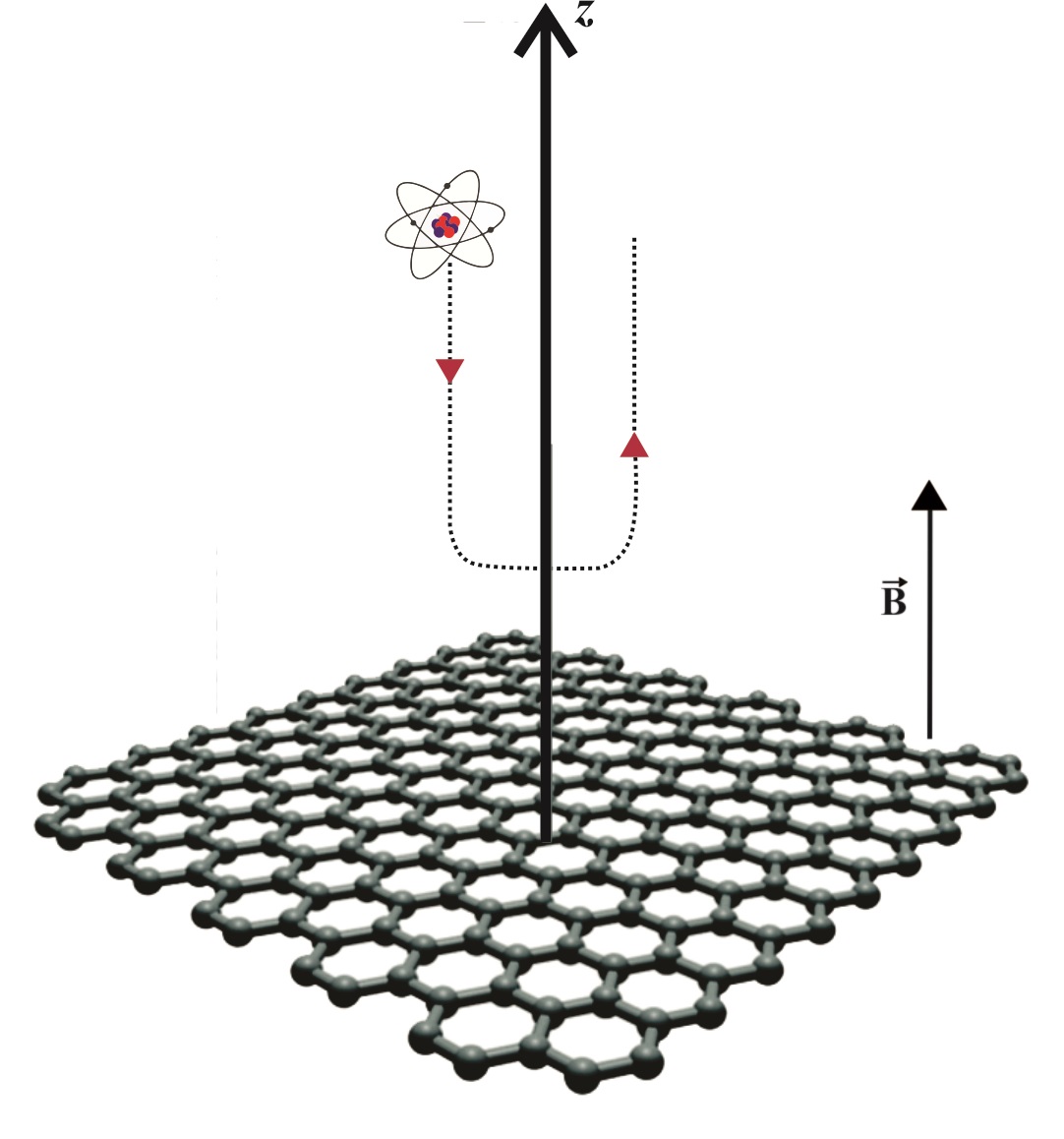}
  \caption{Atom being reflected by the attractive Casimir-Polder force exerted  by a graphene sheet before the atom reaches the sheet. Graphene is under the action of an external  magnetic field perpendicular to the sheet.}
  \label{Figure1}
\end{figure}

We consider a system constituted of a beam of atoms moving towards a graphene sheet with normal incidence, with an applied magnetic field, perpendicular to the sheet as shown in Fig.~\ref{Figure1}. 
The common procedure to calculate the QR  probability $R$ with normal incidence of the atoms towards the wall under consideration is to solve a one-dimensional Schr\"odinger equation with the corresponding  potential $U(z)$ ($z$ being the distance from the atom to the wall) and appropriate boundary conditions. It is usual in the literature \cite{Friedrich-2002, Segev-1997, Oberst-2005,Arnecke-2006} to consider the van der Waals potential $U_{vdw}(z) = - C_3/z^3$, valid for  short distances, or the (asymptotic) retarded potential ${U_{ret}(z) = -C_4/z^4}$, valid for large distances, $C_3$ and $C_4$ being positive constants, or even a simple phenomenological interpolating potential of the form $U_{int}(z) = -C_4/[z^3(z + \ell)]$, where $\ell$ is a parameter with dimensions of length which depends on the incident atom \cite{Judd-2011, Voronin-2005}. A detailed discussion on the comparison of this kind of phenomenological potential and the exact one given by Lifshitz formula can be found in Ref.\cite{Bezerra-2008}. In the present work we do not restrict ourselves to the approximate expressions for the interacting potential between the atom and the wall. Rather, we compute the QR probability by using the full Casimir-Polder potential valid for all distance regimes, which constitutes an important methodological progress with respect to the vast majority of existing theoretical works on QR. In the present case, this potential is the Casimir-Polder potential between a neutral but polarizable atom and a graphene sheet with a magnetic field acting perpendicularly on the sheet at low temperatures, which is given by \cite{Cysne-Kort-Kamp_2014}
\begin{eqnarray}
U(z)&=&\frac{\hbar}{\epsilon_0 c^2}\int_0^{\infty}\frac{d\xi}{2\pi} \xi^2 \,\alpha(i\xi)\int\frac{d^2\textbf{k}}{(2\pi)^2}\frac{e^{-2\kappa z}}{2\kappa}  \cr\cr
&& \hspace{-40pt}\times \Bigg[ r^{s,s}(\textbf{k},i\xi, B)- \Bigg(1+\frac{2c^2k^2}{\xi^2}\Bigg)r^{p,p}(\textbf{k},i\xi, B)\Bigg]\, ,
\label{u}
\end{eqnarray}
where ,
$\kappa=\sqrt{\xi^2/c^2 + k^2}$, $\alpha(i\xi)$ is the electric polarizability of the atom, and $r^{s,s}(\textbf{k},i\xi, B)$,
$r^{p,p}(\textbf{k},i\xi, B)$ are the diagonal reflection coefficients associated to graphene ($s$ and $p$ mean, as usual,  the transverse electric and transverse magnetic polarizations, respectively). The reflection coefficients of graphene in presence of magnetic field, the model for atomic polarizabilities,  and a discussion of the profile of function $U(z)$ can be found in the Appendix.

For an atom of mass $m$ and  energy $E$ under the action of a potential $U(z)$, the Schr\"odinger equation reads
\begin{eqnarray}
\frac{\partial^2 \psi (z)}{\partial z^2}+\frac{p^2(z)}{\hbar^2}\psi(z)=0, \label{SE}
\end{eqnarray}
where
\begin{equation}\label{p}
p(z) = \sqrt{2m [E-U(z)]}\, .
\end{equation}
Since the WKB solutions are good approximations when the atom is far from the graphene sheet (compared to the length scale associated with CP interaction, namely, $c/\xi_l$ [see the table \ref{table1} in the appendix]), it is convenient to try a solution of the form \cite{Berry-1972},

\begin{eqnarray}
\psi(z)=\frac{c_+(z)}{|\sqrt{p(z)}|}e^{i\phi(z)}+\frac{c_-(z)}{\sqrt{|p(z)|}}e^{-i\phi(z)}\, , \label{WKB}
\end{eqnarray}
with $\phi(z)$  given by
\begin{eqnarray}
\phi(z)=\int^{z}_{z_0}\frac{p(z')}{\hbar}dz' \, . \label{WKBphi}
\end{eqnarray} 
Note that there is no approximation in writing previous equations, since  $c_+(z)$ and $c_-(z)$ still need to be determined.  However, the previous ansatz is a very convenient one, since it  transforms the second order Schr\"odinger equation into a set of two coupled first order differential equations for $c_+(z)$ and $c_-(z)$. In fact, substituting Eq(s) (\ref{WKB}) and (\ref{WKBphi}) into Eq.(\ref{SE}) it is straightforward to show that
\begin{eqnarray}\label{DE1}
\frac{\partial c_{+}(z)}{\partial z} &=& e^{- 2 i \phi (z)}\frac{c_-(z)}{2p(z)}\frac{\partial p(z)}{\partial z}\\\cr
\frac{\partial c_-(z)}{\partial z} &=& e^{+2 i \phi (z)}\frac{c_+(z)}{2p(z)}\frac{\partial p(z)}{\partial z}\, .
\label{DE2}
\end{eqnarray}
Reagarding boundary conditions, it is reasonable to impose that $c_+(0) = 0$ and $c_-(0) = 1 $, which means that any atom that reaches the graphene sheet will not be reflected, but rather adsorbed to it. By definition, the quantum reflection probability is \cite{Dufour-2013},
\begin{eqnarray}
R=\lim_{z\rightarrow \infty}\Big| \frac{c_+(z)}{c_-(z)}\Big|^2\, . \label{QRP}
\end{eqnarray}
The previous set of coupled first order differential equations for $c_+(z)$ and $c_-(z)$ will be solved numerically. Information about the  efficiency of a certain potential $U(z)$  to give rise to QR can be extracted from the  function \cite{Dufour-2013,Dufour-2-2013},
\begin{eqnarray}
Q(z)=\frac{\hbar^2}{2p^2(z)}\Bigg[ \frac{\phi '''(z)}{\phi'(z)}-\frac{3}{2}\Bigg(\frac{\phi'' (z)}{\phi'(z)}\Bigg)^2\Bigg]\, , \label{QF}
\end{eqnarray}
with $p(z)$ and $\phi(z)$ given by (\ref{p}) and (\ref{WKBphi}). It can be shown that the highest  probabilities  of occurrence of QR correspond to the regions of highest values of $Q(z)$ \cite{Dufour-2013, Dufour-2-2013, Dufour-2015-tesis}. Moreover, for a given energy, this function exhibits a peak. Let us denote by $z_m$ the position of this peak, that is, the distance between the atom and the graphene sheet where $Q(z)$ is maximum. In order to solve numerically the set of coupled differential equations (\ref{DE1}) and (\ref{DE2}), we choose a point close to graphene sheet ($z_i$) and a point far from graphene ($z_f$) such that $z_i \ll z_m \ll z_f$. The differential equations are solved in the region between $z_i$ and $z_f$ which contains the peak of  $Q(z)$. The limit in Eq. (\ref{QRP}) is  numerically calculated by taking values of  coefficients $c_+(z)$ and $c_-(z)$ at point $z_f$. Parameters $z_f$ and $z_i$  are convergence parameters. We have numerically established that a good convergence of the results occurs whenever changes in $z_i$ and $z_f$ do not affect any more the value of the QR probability $R$. A detailed discussion of this method can be found in reference \cite{Dufour-2015-tesis}.
 
As already mentioned, in most calculations of QR, an interpolation between the non-retarded van der Waals potential ($\approx - C_3/z^3$) and the retarded potential ($\approx - C_4/z^4$) regimes is used, allowing for a semi-analytical solution of the coupled differential equations for the WKB coefficients $c_+(z)$ and $c_-(z)$. In the present work we solve the coupled equations (\ref{DE1}) and  (\ref{DE2}) by  using the complete Casimir-Polder potential whose dependence with distance is quite involved, so that a full numerical procedure is demanded. The consideration of the whole potential is crucial for obtaining reliable results on the control of QR on graphene with the aid of an external magnetic field, and constitutes an important methodological advance with respect to the traditional theoretical approach to QR. In the next section we present the main results  of this work obtained by numerical calculations. As we shall see, the complex behavior of the complete Casimir-Polder potential with the external magnetic field has non trivial consequences on the QR of different atomic specimens. It is worth mentioning that the Zeeman coupling in graphene can be safely neglected due to the more pronounced effect of Landau quantization of electronic motion, related to the relativistic nature of the Dirac spectrum \cite{Gusynin-2005-1, Gusynin-2005-2}. In addition, we have verified that the Zeeman effect upon the atoms only leads to negligible corrections in the computation of the dispersive energy interaction between graphene and the atoms, for all species considered, as previously shown in Ref.~\cite{Cysne-Kort-Kamp_2014}.

%-----------------------------------------------------------
\section{Results and discussions}

In order to investigate the consequences of the application of a magnetic field perpendicular to graphene sheet in the probability of an atom to be reflected by this sheet, we consider three atomic species of experimental relevance: He, Na and Rb. These atomic species have already been used in QR experiments\cite{Nayak-1983, Berkhout-1989, Doyle-1991, Yu-1993, Pasquini-2004, Pasquini-2006, Marchant-2016}. The atomic polarizabilities of these atoms will be used in the Casimir-Polder potential [Eq. (\ref{u})] and are described by the single Lorentz-oscillator model, whose fitted parameters are well known from the literature \cite{Khusnutdinov-2016} [see Appendix]. Once we would like our results to be valid regardless the retard/non-retard regime, we numerically evaluate the Casimir-Polder potential in the distance range $z_i< z< z_f$ at given values of the magnetic field [see Appendix]. In order to solve numerically the system of coupled  differential equations (\ref{DE1}) and (\ref{DE2}), we use a standard procedure of interpolation by a polynomial function of successive points of the CP potential and, finally,  we may obtain the QR reflection probability $R$ using Eq. (\ref{QRP}).

\begin{figure}[h]
\vspace{0.1in}
  \centering
  \includegraphics[scale = 0.55]{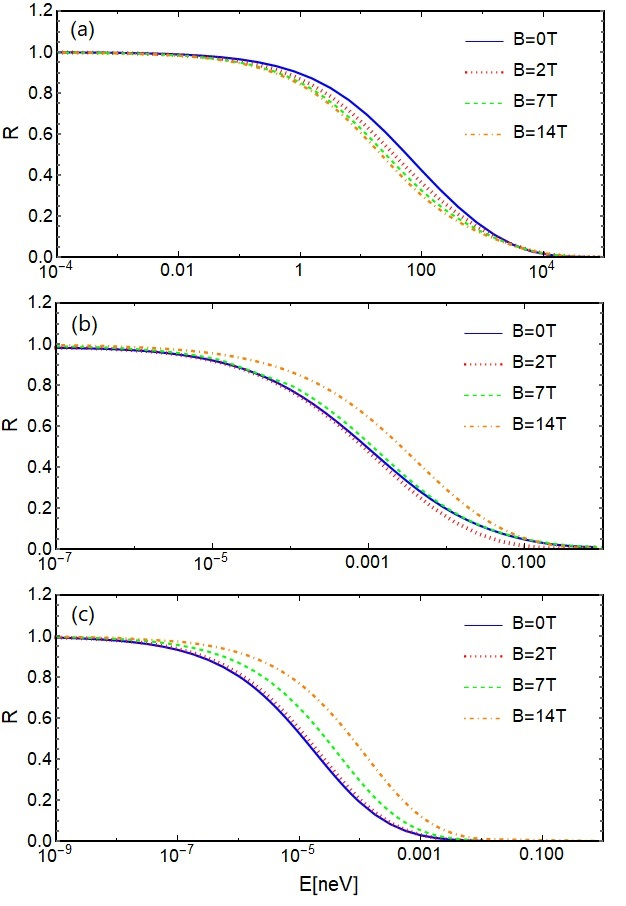}
  \caption{Quantum reflection probability $R$ as a function of the energy of the incident particle for (a) He, (b) Na and (c) Rb at given values of the external magnetic field. 
  We considered the chemical potential of the graphene sheet fixed at $\mu_c=0.115$eV.}
  \label{Figure2}
\end{figure}

In Fig. \ref{Figure2} we plot the QR probabilities $R$ of He, Na and Rb as functions of  their incident energies for four values of  the magnetic field ($B=0,2,7,14$T). The chemical potential of graphene sheet is set to $\mu_c=0.115$eV. Note that, in all cases, $R\rightarrow 1$ as $E \rightarrow 0$, which is a direct consequence of the fact that the de Broglie wavelength associated to the atom increases as $E$ decreases, i.e. the wave-like nature of the particles becomes more important the lower their energy is. By the same token, $R\rightarrow 0$ as $E \rightarrow \infty$, as expected, since  the de Broglie wavelength decreases as $E$ increases. The limit where $\lambda \rightarrow 0$ is analogous to the geometrical optics limit in wave optics and hence the particle behavior of the atom must show up. In between these two regimes, there is an intermediate one in which QR is substantially influenced  by the external magnetic field. This is one of the main results of our paper, showing that a magnetic field applied to a graphene sheet may be used as an efficiently way of controlling and tuning QR by an external agent.  Moreover,  this influence depends on the  atom under consideration. For He atom [Fig. \ref{Figure2}, panel (a)] the QR probability at a given energy decreases when the magnetic field  is applied ($B=2, 7, 14$T curves). On other hand, for Rb atom [Fig. \ref{Figure2}, panel (c)] the QR probability is enhanced in presence of magnetic field ($B=2,7,14$T). For Na [panel (b) of Fig. \ref{Figure2}], there is a non-monotonic behaviour with B, i. e. QR decreases for $B=2$T and  increases for $B=7, 14$T when compared to the case of $B=0$ T. 

\begin{figure}[h]
\vspace{0.1in}
  \centering
  \includegraphics[scale = 0.55]{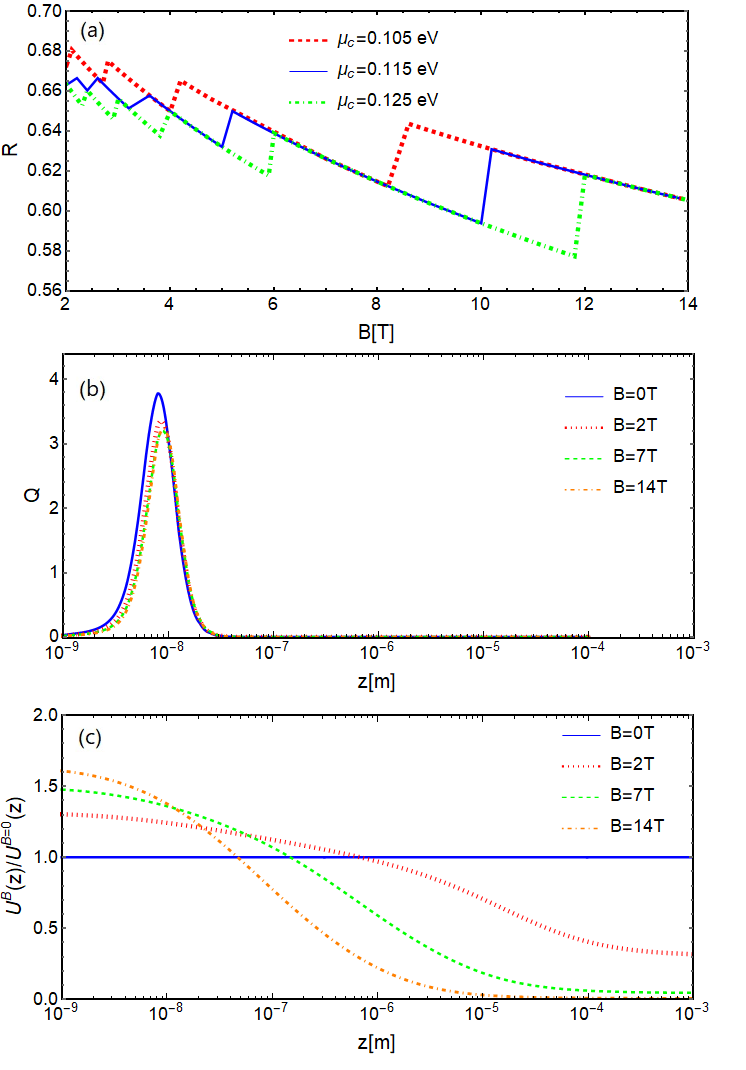}
  \caption{For He atom, Panel (a): QR probability as a function of the external magnetic field for chemical potentials of the graphene sheet $\mu_c=0.105$ eV (Red Dashed), $\mu_c=0.115$ eV (Blue Solid), $\mu_c=0.125$ eV (Green Dot-Dashed). Panel (b): Function $Q(z)$ of Eq. (\ref{QF}) for chemical potential of the graphene sheet $\mu_c=0.115$ eV and external magnetic field intensities $B=0, 2, 7, 14$ T. For all data at the panels (a) and (b) we set the energy of the incident particle $E=10^1$ neV. Panel (c): Relative variation of CP energy with the magnetic field as a function of distance ($U^B(z)/U^{B=0}(z)$) for the chemical potential of the graphene sheet $\mu_c=0.115$eV external magnetic field intensities $B=0, 2, 7, 14$ T.}
  \label{Figure3}
\end{figure}

\begin{figure}[h]
\vspace{0.1in}
  \centering
  \includegraphics[scale = 0.55]{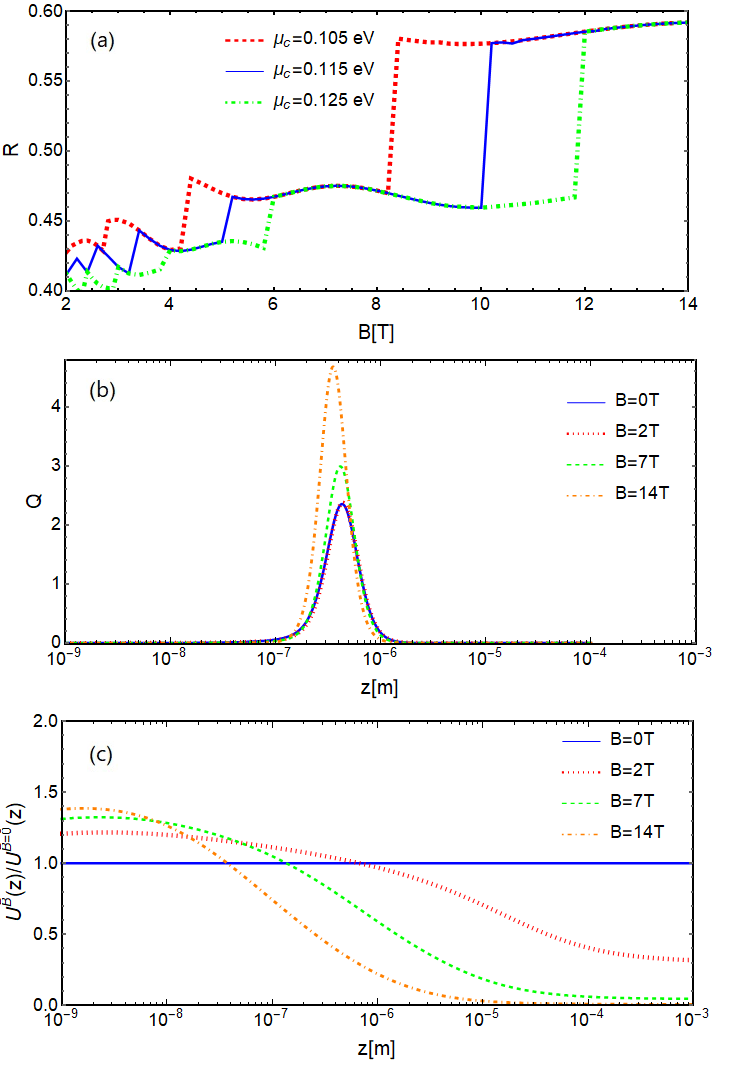}
  \caption{For Na atom, Panel (a): QR probability as a function of the external magnetic field for chemical potentials of the graphene sheet $\mu_c=0.105$ eV (Red Dashed), $\mu_c=0.115$ eV (Blue Solid), $\mu_c=0.125$ eV (Green Dot-Dashed). Panel (b):  Function $Q(z)$ of Eq. (\ref{QF}) for chemical potential of the graphene sheet $\mu_c=0.115$ eV and external magnetic field intensities $B=0, 2, 7, 14$ T. For all data at the panels (a) and (b) we set the energy of the incident particle $E=10^{-3}$ neV. Panel (c): Relative variation of CP energy with the magnetic field as a function of distance ($U^B(z)/U^{B=0}(z)$) for the chemical potential of the graphene sheet $\mu_c=0.115$eV external magnetic field intensities $B=0, 2, 7, 14$ T.}
  \label{Figure4}
\end{figure}

\begin{figure}[h]
\vspace{0.1in}
  \centering
  \includegraphics[scale = 0.55]{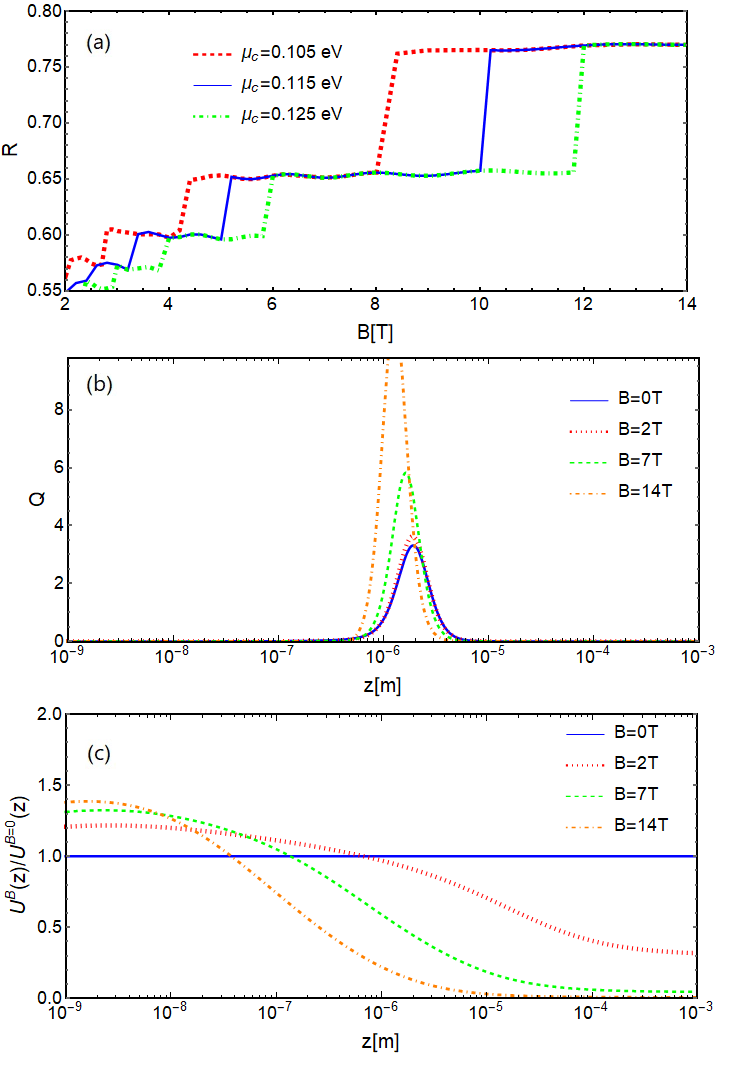}
  \caption{For Rb atom, Panel (a): QR probability as a function of the external magnetic field for chemical potentials of the graphene sheet $\mu_c=0.105$ eV (Red Dashed), $\mu_c=0.115$ eV (Blue Solid), $\mu_c=0.125$ eV (Green Dot-Dashed). Panel (b):  Function $Q(z)$ of Eq. (\ref{QF}) for chemical potential of the graphene sheet $\mu_c=0.115$ eV and external magnetic field intensities $B=0, 2, 7, 14$ T. For all data at the panels (a) and (b) we set the energy of the incident particle $E=10^{-5}$ neV. Panel (c): Relative variation of CP energy with the magnetic field as a function of distance ($U^B(z)/U^{B=0}(z)$) for the chemical potential of the graphene sheet $\mu_c=0.115$eV and external magnetic field intensities $B=0, 2, 7, 14$ T.}
  \label{Figure5}
\end{figure}

In order to understand these different behaviors of the QR probability for different atoms, it is necessary to analyze the interplay between the tunability of CP energy with the magnetic field at different distance regimes and the function $Q(z)$. The position of the peak of $Q(z)$ indicates the most probable region of space where QR can occur. To this end, we show in the Figs. \ref{Figure3}, \ref{Figure4}, and \ref{Figure5} the dependence of QR probability [panel (a)], $Q(z)$ function [panel (b)] and the relative change of CP energy with magnetic field  [panel (c)] for He, Na and Rb atoms, respectively.  In the case of He atom, we show in panel (a) of Fig. \ref{Figure3} the quantum reflection probability as a function of the applied magnetic field for three different values of the chemical potential of the graphene sheet. The incident energy of the atom was fixed at a value that leads to optimal tunability in a magnetic field, as it can be seen in Fig. \ref{Figure2} ($E=10^1$neV for the case of He in Fig. \ref{Figure3}, panels (a) and (b)). The QR probability as a function of the magnetic field for He atom shows a general decrescent behavior, consistent with the discussion of the previous paragraph. In addition to this decrescent trend, there exist discontinuities that are related to discontinuities in the CP energy whenever a Landau level of the spectrum of graphene (which varies with B) crosses the chemical potential \cite{Cysne-Kort-Kamp_2014}. The effect of varying the chemical potential is just a shift in values of the magnetic field where these discontinuities take place. In panel (b) of Fig. \ref{Figure3}, we show the function $Q(z)$ for chemical potential $\mu_c = 0.115$ eV, incident energy of He atom $E=10^1$ neV and four values of the external magnetic field. The maximum of $Q(z)$ function occurs at distances from the graphene sheet where the CP energy is enhanced by the external magnetic field [As it can be seen from the panel (c) of Fig. \ref{Figure3}]. This enhancement of CP energy with the magnetic field in the region of space where QR takes place is related to the decrescent behavior of the QR probability with the magnetic field for the He atom discussed in the previous paragraph. In the case of Rb, in the panel (a) of the Fig. \ref{Figure5}, we show the QR probability as a function of the external magnetic field for three values of the chemical potential.  The incident energy of the Rb atom was set to correspond to the optimal tunability in a magnetic field on Fig. \ref{Figure2} ,{\it i. e.}, $E=10^{-5}$neV for panels (a) and (b) in Fig. \ref{Figure5}. Again, the QR probability as a function of the magnetic field shows discontinuities associated with the crossing of Landau levels through the chemical potential in the electronic spectrum of the graphene sheet. However, here QR probability has a crescent behavior. In the panel (b) of the figure, it is possible to see that the maximum of the function $Q(z)$ occurs in the region of space where the CP energy decreases with the presence of the magnetic field. This explains the different behaviors of QR probability with magnetic field for different atoms. In the case of Rb atom, QR takes place in a region where the CP energy decreases with the magnetic field, so that the QR probability is enhanced. Furthermore, in between the discontinuities at QR probability for Rb [panel (a) Fig. \ref{Figure5}], there exist plateaus. This behavior is a direct consequence of the plateaus in the CP energy as a function of the magnetic field at long-distance regimes (where QR takes place), which occur due the prevalence of low-frequency modes in Lifshitz formula \cite{Cysne-Kort-Kamp_2014}. Finally, in the case of Na (shown in Fig. \ref{Figure4}), the QR takes place at intermediate distances, i. e., in the region between the short distance regime, where CP energy increases with the magnetic field, and the long-distance regime, where CP energy decreases with the magnetic field. This explains the non-monotonic dependence of QR probability with the magnetic field for Na atom discussed in the previous paragraph and shown in the panel (b) of Fig.\ref{Figure2}.

All results presented in this paper have been derived by the  Casimir-Polder interaction computed from Lifshitz formula at zero temperature  [Eq. (\ref{u})]. The detailed study of the effect of finite temperature in quantum reflection is beyond the scope of the present work, but we shall briefly comment on the thermal effects on our results. It is well known that thermal effects in dispersive forces, which have been investigated in {\it e.g. } Ref. ~\cite{Bhuman-Schell}, change the long-distance regime of the Casimir-Polder interaction. This distance regime is characterized by the thermal wavelength, which is proportional to $(k_B T)^{-1}$ and thermal corrections may influence the quantum reflection quantitatively \cite{Bezerra-2008}. These corrections should be more pronounced for heavy atoms, for which quantum reflection is dominated by long-distance regimes of the Casimir-Polder potential (see Appendix).  Moreover, finite temperature effects cause a broadening in the Fermi-Dirac distribution which, in turn, makes the discontinuities in the Casimir-Polder potential as a function of the magnetic field smear out. However, thermal effects do not prevent QR to be tuned by an magnetic field, since the Casimir-Polder potential at finite temperature is still substantially changed by a magnetic field \cite{Cysne-Kort-Kamp_2014}. Therefore, we expect that even at high temperatures, quantum reflection by a graphene sheet  may still be tuned by a magnetic field, although the discontinuities  present in panels (a) of Figs. \ref{Figure3},\ref{Figure4}, and \ref{Figure5}  will disappear. The discontinuities in Figs \ref{Figure3}, \ref{Figure4} and \ref{Figure5} must be observable for small temperatures (thermal energy $k_B T$) compared to the energy spacing between two successive  Landau levels. These energy spacings become larger for high magnetic fields. This fact  can be used in order to reach the regime where discontinuities in quantum reflection probability should be observable.

%-----------------------------------------------------------
\section{Conclusions}

In summary, we have demonstrated that an external magnetic field may be used to tune Quantum Reflection (QR) in a graphene sheet. We have calculated the attractive Casimir-Polder potential using Lifshitz formula, valid at any distance between the atom and the graphene sheet. We have considered three atomic species of experimental relevance He, Na, Rb, for which we calculate the QR probability using a full numerical approach. We identify three distinct different behaviors in the QR probability, depending on the magnitude of the applied magnetic field, which can be explained in terms of the different characteristic distance regimes of the Casimir-Polder potential. We also conclude that the effects of the magnetic field on QR are more pronounced for lighter atomic species at the short distance regime of the Casimir-Polder, while for heavier atoms the magnetic effects tend to be more intense in the long-distance regime. However, in all cases and for all the investigated atoms, QR in graphene can be efficiency tuned by applying an external magnetic field. We show that the QR probability exhibits discontinuities as a result of the quantization of the electronic spectrum of graphene. These discontinuities persist at low temperatures and for high values of the magnetic field. Altogether our findings not only allow for an alternative way to control quantum reflection at the nanoscale, but also open the door for the design of novel reflective optical elements, such as Fresnel mirrors~\cite{judd2010}, which can be used for tunable reflective focusing of matter waves.

\section*{Acknowledgements}
The authors thank the Brazilian Agencies CNPq, CAPES, and FAPERJ for their financial support.

%%%% %%%%%%  APENDICE   %%%%%%%%%%%

\appendix
\section{Reflection coefficients, atomic polarizabilities, and Casimir-Polder energy in presence of magnetic field.\label{append}}

The reflection coefficients of graphene sheet under the influence of a magnetic field applied perpendicular to the sheet can be obtained by using Maxwell's equations with the appropriate boundary conditions,\cite{Macdonald_PRL, Kort-Kamp-Amorim_2015, Sounas_2012}
\begin{eqnarray}
&&r^{s,s}({\bf k},i\xi,B) = \nonumber \\
&& \frac{2\sigma_{xx}(i\xi, B) Z^h +
\eta_0^2[\sigma_{xx}(i\xi, B)^2+\sigma_{xy}(i\xi, B)^2]}
{-\Delta({\bf k},i\xi, B)}\, ,  
\end{eqnarray}
\begin{eqnarray}
&&  r^{p,p}({\bf k},i\xi,B)= \nonumber \\
&&\dfrac{2\sigma_{xx}(i\xi, B) Z^e +
\eta_0^2[\sigma_{xx}(i\xi, B)^2+\sigma_{xy}(i\xi, B)^2]}
{\Delta({\bf k},i\xi, B)}\, ,  
\end{eqnarray}
\begin{eqnarray}
&&\Delta({\bf k},i\xi, B) = [2 + Z^h \sigma_{xx}(i\xi, B)][2
+ Z^e \sigma_{xx}(i\xi,B)] \nonumber \\
&&  +[\eta_0\sigma_{xy}(i\xi, B)]^2  \, ,
\label{RefCoefs}
\end{eqnarray}
where $Z^h = \xi \mu_0 / \kappa$, $Z^e = \kappa/(\xi \epsilon_0)$, and
$\eta_0^2=\mu_0/\epsilon_0$. Besides, $\sigma_{xx}(i\xi, B)$ and
$\sigma_{xy}(i\xi, B)$ are the longitudinal and transverse conductivities of
graphene, respectively. The electric conductivity tensor of graphene under an external magnetic field is
well known and reads \cite{Gusynin1, Gusynin2}
\begin{widetext}
\vskip -0.4cm
\begin{eqnarray}
\sigma_{xx}(i\xi,B) &=& \dfrac{e^3v_F^2 B\hbar(\xi+\tau^{-1})}{\pi} \sum\limits_{n=0}^\infty \Bigg\{\dfrac{n_F(M_n)-n_F(M_{n+1})+n_F(-M_{n+1})-n_F(-M_n)}    {D_n(M_{n+1}-M_n)}
+(M_n\to-M_n)\Bigg\}\, , \label{Conductivity1}\\
\sigma_{xy}(i\xi,B)&=&\dfrac{e^3v_F^2B}{-\pi}\sum\limits_{n=0}^\infty \{n_F(M_n)-n_F(M_{n+1})-n_F(-M_{n+1})+n_F(-M_n)\} \Bigg[\dfrac{1}{D_n}
+(M_n\to-M_n) \Bigg]\, ,
\label{Conductivity2}
\end{eqnarray}
\end{widetext}
where $1/\tau$ is a phenomenological scattering rate which causes a small broadening in the Landau levels, $n_F(E)= \Theta (\mu_c-E)$ is the
Fermi-Dirac distribution at zero temperature, $D_n = (M_{n+1}-M_n)^2+\hbar^2(\xi+\tau^{-1})^2$,
$M_n=\sqrt {n} M_1$ are the Landau energy levels, $M_1^2 = 2 \hbar e B v_F^2$
is the Landau energy scale and $v_F \simeq 10^6$ m/s is the Fermi velocity.
The scattering time can be obtained by the fit of the Drude model to optical conductivities given by experiments or numerical calculations\cite{Cysne-2016}. In the present paper, we set $\tau = 1.84 \times 10^{-13}$ s, which is a characteristic value for weakly disordered graphene samples. $\mu_c$ is the chemical potential of the graphene sheet. 

The atomic polarizabilities of the different atoms can be modeled by a single Lorentz-oscillator model. Refinements of this model do exist \cite{Khusnutdinov-2016} but they do not modify the main conclusions of the present paper concerning QR. The atomic polarizability in the imaginary frequency axis is given by
\begin{eqnarray}
\alpha_l(i\xi)=\frac{\alpha_l(0)}{1+\frac{\xi^2}{\xi_l^2}}.
\label{alpha}
\end{eqnarray}
The parameters of Eq. (\ref{alpha}) for each atom considered are given in the table \ref{table1} ($1 Au= 1.648 \times 10^{-41} C^2m^2J^{-1}$). With these coefficients, one can compute the Casimir-Polder energy between an atom and graphene sheet using Eq. (\ref{u}).

\begin{table}[h!]
\centering
\begin{tabular}{||c c c||} 
 \hline
 $l$ & $\alpha_l(0) (Au)$ & $\xi_l (eV)$ \\ [0.5ex] 
 \hline\hline
 He & 1.384 & 27.64 \\
 Na & 162.6 & 2.13  \\
 Rb & 318.6 & 1.68 \\ [1ex] 
 \hline
\end{tabular}
\caption{Parameters of the single Lorentz-oscillator model of Eq. (\ref{alpha}), for He, Na and Rb atoms \cite{Khusnutdinov-2016}.}
\label{table1}
\end{table}

\begin{figure}[h]
\vspace{0.1in}
  \centering
  \includegraphics[scale = 0.55]{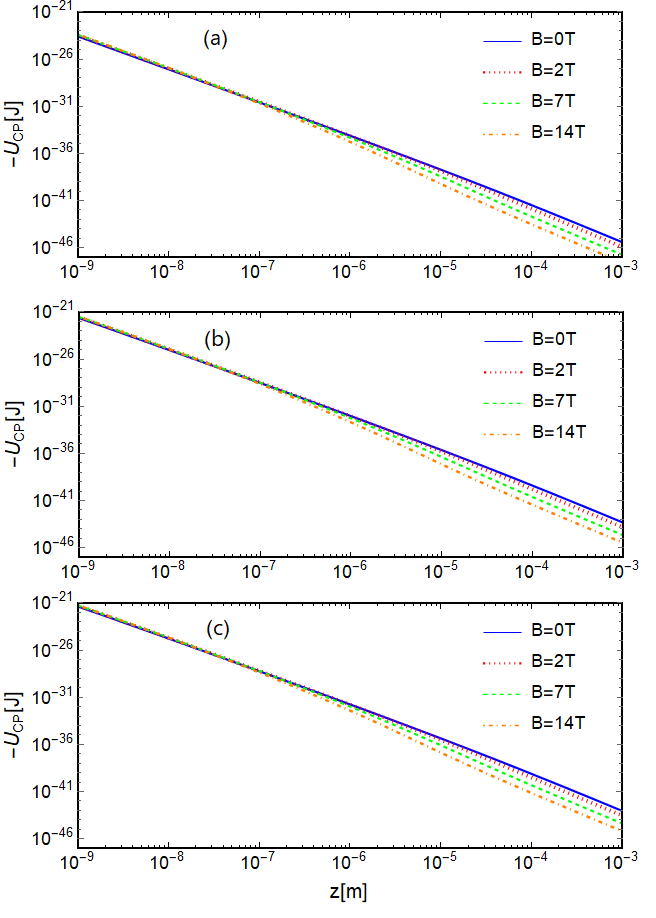}
  \caption{Casimir-Polder energy as function of distance between graphene and atom for different values of external magnetic field: (a) He (b) Na (c) Rb. We set the chemical potential of graphene sheet $\mu_c=0.115$eV in this figure.}
  \label{FigureA1}
\end{figure}

We show in Fig. \ref{FigureA1} the Casimir-Polder energy computed by using Eq. (\ref{u}) for the  three atomic species considered in this paper [(a) He, (b) Na, (c) Rb]. The polarizability of each atom was modeled with a single Lorentz-oscillator model \cite{Khusnutdinov-2016}. For the long-distance (retarded) regime ($z\gg10^{-6}$m), the function $U(z)$ has the known asymptotic behaviour $C_4/z^4$ for the three atoms. The $C_{4}$-coefficient is a function of the intensity of the applied magnetic field ($B$) and of the chemical potential of the graphene sheet ($\mu_c$). It is clear from Fig. \ref{FigureA1} that the magnetic field decreases the intensity of the Casimir-Polder potential in the long-distance regime for the three atoms. On the other hand, for short distances ($z\ll 10^{-6}$m), the Casimir-Polder  potential  does not obey a universal asymptotic power-law dependence for interaction with graphene in the presence of a magnetic field. Moreover, we have numerically verified that the Casimir-Polder potential is enhanced due the presence of the magnetic field in the short-distance regime. The enhancement of CP energy at the short-distance regime is more evident when we plot the ratio $U^B(z)/U^{B=0}(z)$, as was done in panel (c) of Figs. \ref{Figure3} (He), \ref{Figure4} (Na), \ref{Figure5} (Rb).

\newpage
%------------------------------------------------------------------------------------------------%


\begin{thebibliography}{99}

\bibitem{DeBroglie} L. de Broglie, Nature {\bf 112}, 540 (1923).

%\bibitem{Schrodinger} E. Schr\"odinger, Phys. Rev. {\bf 28}, 1049 (1926).


\bibitem{Brekhovskikh-2012} L. Brekhovskikh, Waves in layered media, 2nd (Elsevier, 2012) (cited on p. 7).

%H-He reflected by liquid He
\bibitem{Nayak-1983} V.U. Nayak, D.O. Edwards, and N. Masuhara, Phys. Rev. Lett. {\bf 50}, 990 (1983).
\bibitem{Berkhout-1989} J. J. Berkhout et al., Phys. Rev. Lett. {\bf  63}, 1689 (1989).
\bibitem{Doyle-1991} J.M. Doyle et al., Phys. Rev. Lett. {\bf 67}, 603 (1991).
\bibitem{Yu-1993} I. A. Yu et al., Phys. Rev. Lett. 71, 1589 (1993).

%QR of Na atoms
\bibitem{Pasquini-2004} T. A. Pasquini, Y. Shin, C. Sanner, M. Saba, A. Schirotzek, D. E. Pritchard, and W. Ketterle, Phys. Rev.Lett. {\bf 93}, 223201 (2004).
\bibitem{Pasquini-2006} T. A. Pasquini, M. Saba, G.-B. Jo, Y. Shin, W. Ketterle, D. E. Pritchard, T. A. Savas, and N. Mulders, Phys. Rev. Lett. {\bf 97}, 093201 (2006). 

%QR of Rb atoms
\bibitem{Marchant-2016} A. L. Marchant, T. P. Billam, M. M. H. Yu, A. Rakonjac, J. L. Helm, J. Polo, C. Weiss, S. A. Gardiner, and S. L. Cornish, Phys. Rev. A {\bf 93}, 021604(R) (2016).

%QR- Oblique incidence
\bibitem{Druzhinina-2003} V. Druzhinina and M. DeKieviet, Phys. Rev. Lett. {\bf 91}, 193202 (2003).
\bibitem{Zhao-2008} B. S. Zhao, S. A. Schulz,  S. A. Meek, G. Meijer, W. Schöllkopf, Phys. Rev. A {\bf 78},010902(R) (2008).

%Study of CP-with QR
\bibitem{Shimizu} F. Shimizu, Phys. Rev. Lett. {\bf 86}, 987 (2001).

%Probing CP by QR
%\bibitem{Bender-2010} H. Bender, Ph. W. Courteille, C. Marzok, C. Zimmermann, and S. Slama, Phys. Rev. Lett. {\bf 104}, 083201 (2010).
\bibitem{Oberst-2005} H. Oberst, Y. Tashiro, K. Shimizu, and F. Shimizu, Phys. Rev. A {\bf 71}, 052901 (2005).
\bibitem{Zhao-2010} B. S. Zhao, H. Christian Schewe, Gerard Meijer, and Wieland Sch\"{o}llkopf, Phys. Rev. Lett. {\bf 105}, 133203 (2010).
\bibitem{Zhao-2013} B. S. Zhao, W. Zhang, and W. Schöllkopf, Mol. Phys. {\bf 111}, 1772 (2013).
\bibitem{Barnea-2017} A. R. Barnea, B. A. Stickler, O. Cheshnovsky, K. Hornberger, and Uzi Even, Phys. Rev. A {\bf 95}, 043639 (2017).
%atomic mirror
\bibitem{Cote-2003} R. C\^{o}t\'{e}, and B. Segev Phys. Rev. A {\bf 67}, 041604(R) (2003).

%atomic mirror pot c3/z3
\bibitem{Segev-1997} B. Segev, R. C\^{o}t\'{e}, and M. G. Raize, Phys. Rev. A 56, R3350(R) (1997).

%Mirotrap atoms
\bibitem{Hammes-2002} M. Hammes, D. Rychtarik, H.-C. Nägerl, and R. Grimm, Phys. Rev. A {\bf 66}, 051401(R) (2002).

%Trap cold atoms with levitating states
\bibitem{Crepin-2017} P.-P. Cr\'{e}pin, G. Dufour, R. Guérout, A. Lambrecht, and S. Reynaud, Phys. Rev. A {\bf 95}, 032501 (2017).
\bibitem{Jurisch-2008} A. Jurisch and J.-M. Rost Phys. Rev. A {\bf 77}, 043603 (2008).

%Atom-optics
\bibitem{Cronin-RMP-2009} A. D. Cronin, J. Schmiedmayer, and D. E. Pritchard, Rev. Mod. Phys. {\bf 81}, 1051 (2009).
\bibitem{Deutschmann-1993} R. Deutschmann , W. Ertmer, and H. Wallis, Phys. Rev. A {\bf 47}, 2169 (1993).
\bibitem{Landragin-1997} A. Landragin, L. Cognet, G. Z. K. Horvath, C. I. Westbrook, N. Westbrook, and A. Aspect Europhys. Lett. {\bf 39}, 485 (1997).
\bibitem{Shimizu-Fujita-2002} F. Shimizu, and J. Fujita, Phys. Rev. Lett. {\bf 88}, 123201 (2002).
\bibitem{Kohno-2003} T. Kohno, F. Shimizu, J. Fujita, and K. Shimizu J. Phys. Soc. Jpn. {\bf 72}, 461 (2003).
\bibitem{Savalli-2002} V. Savalli, D. Stevens, J. Esteve, P. D. Featonby, V. Josse, N. Westbrook, C. I. Westbrook, and A. Aspect  Phys. Rev. Lett. {\bf 88}, 250404 (2002)

%GBAR colaboration-Quantum Reflection
\bibitem{Perez-2015} P. Perez {\it et. al.} (GBAR Collaboration), Hyperfine Interact. {\bf 233}, 21 (2015).
\bibitem{Dufour-2014} G. Dufour et al. The Europ. Phys. J. C {\bf 74} ,2731 (2014). 

%Tunability graphene
\bibitem{Cysne-2016} T. P. Cysne, T. G. Rappoport, Aires Ferreira, J. M. Viana Parente Lopes, and N. M. R. Peres, Phys. Rev. B {\bf 94}, 235405 (2016).
\bibitem{Bordag-2016} M. Bordag, I. Fialkovskiy, and D. Vassilevich, Phys. Rev. B {\bf 93}, 075414 (2016).
\bibitem{Henkel-2018} C. Henkel, G. L. Klimchitskaya, and V. M. Mostepanenko, Phys. Rev. A {\bf 97}, 032504 (2018).
\bibitem{Cysne-Kort-Kamp_2014} T. Cysne, W. J. M. Kort-Kamp, D. Oliver, F. A. Pinheiro, F. S. S. Rosa, and C. Farina, Phys. Rev. A {\bf 90}, 052511 (2014).
\bibitem{Macdonald_PRL} W.-K. Tse and A. H. MacDonald, Phys. Rev. Lett. {\bf 109}, 236806 (2012).
\bibitem{Nichols-2016} N. S. Nichols, A. Del Maestro, C. Wexler, and V. N. Kotov, Phys. Rev. B {\bf 93}, 205412 (2016).
\bibitem{Khusnutdinov-2016} Nail Khusnutdinov, Rashid Kashapov, and Lilia M. Woods, Phys. Rev. A {\bf 94}, 012513 (2016).
\bibitem{Khusnutdinov-2018} N. Khusnutdinov, R. Kashapov, and L. M. Woods, 2D Materials {\bf 5}, 035032 (2018).

%QR-Graphene
\bibitem{Judd-2011} T. E. Judd, R. G. Scott, A. M. Martin, B. Kaczmarek, and T. M. Fromhold, New J. Phys. {\bf 13}, 083020 (2011).

%Livro Miloni
\bibitem{Milonni} P. W. Milonni, {\it The Quantum Vacuum: An Introduction to Electrodynamics}, (Academic Press, San Diego, CA, 1994).

\bibitem{Lifshitz-1956} E. M. Lifshitz, Soviet Phys. JETP {\bf 2}, 73 (1956).

%Neclecting Zeeman
\bibitem{Gusynin-2005-1} V. P. Gusynin and S. G. Sharapov ,Phys. Rev. B {\bf 71}, 125124 (2005).
\bibitem{Gusynin-2005-2} V. P. Gusynin and S. G. Sharapov, Phys. Rev. Lett. {\bf 95}, 146801 (2005).

%QR by different regimes of CP
\bibitem{Friedrich-2002} H. Friedrich, G. Jacoby, and C. G. Meister, Phys. Rev. A {\bf 65}, 032902 (2002).

%potential -1/r4 -1/r3
\bibitem{Arnecke-2006} F. Arnecke, H. Friedrich, and J. Madro\~{n}ero, Phys. Rev. A {\bf 74}, 062702 (2006).
\bibitem{Voronin-2005} A. Y. Voronin,P. Froelich,and B. Zygelman, Phys. Rev. A {\bf 72}, 062903 (2005).

%Thermal corrections to QR.
\bibitem{Bezerra-2008} V. B. Bezerra, G. L. Klimchitskaya, V. M. Mostepanenko, and C. Romero, Phys. Rev. A {\bf 78}, 042901 (2008).

%WKB-solution
\bibitem{Berry-1972} M. Berry and K. Mount, Rep. Prog. Phys. {\bf 35}, 315 (1972). 

\bibitem{Dufour-2013} G. Dufour, A. G\'erardin, R. Gu\'erout, A. Lambrecht, V. V. Nesvizhevsky, S. Reynaud, and A. Yu. Voronin, Phys. Rev. A {\bf 87}, 012901 (2013).

\bibitem{Dufour-2-2013} G. Dufour, R. Guérout, A. Lambrecht, V. V. Nesvizhevsky, S. Reynaud, and A. Yu. Voronin, Phys. Rev. A {\bf87}, 022506 (2013).

\bibitem{Dufour-2015-tesis} G. Dufour. {\it Quantum reflection from the Casimir-Polder potential}. Quantum Physics [quantph]. Université Pierre et Marie Curie - Paris VI, (2015).

\bibitem{judd2010} T. E. Judd, R. G. Scott, G. Sinuco, T. W. A. Montgomery,
A. M. Martin, P. Kruger, and T. M. Fromhold, New J. Phys.{\bf 12}, 063033 (2010).

%Citar
%\bibitem{Buhmann-2018} S. Y. Buhmann, V. N. Marachevsky, and S. Scheel, Phys. Rev. A {\bf 98}, 022510 (2018).
%\bibitem{Fuchs-2017} S. Fuchs, F. Lindel, R. V. Krems, G. W. Hanson, M. Antezza, and S. Y. Buhmann, Phys. Rev. A {\bf 96}, 062505 (2017).
%\bibitem{Rodrigues-Lopez-2018}  P. Rodriguez-Lopez, W. J. M. Kort-Kamp, D. A. R. Dalvit, L. M. Woods, Nat. Comm. {\bf 8}, 14699 (2017).
%\bibitem{Belen-2018} M. Bel\'{e}n Farias, W. J. M. Kort-Kamp, and D. A. R. Dalvit, Phys. Rev. B {\bf 97}, 161407(R) (2018).
\bibitem{Bhuman-Schell} S. Y. Buhmann and S. Scheel Phys. Rev. Lett. {\bf 100}, 253201 (2008).
%Mesoscopic model for reflection coeficients in presence of magnetic field
\bibitem{Kort-Kamp-Amorim_2015} W. J. M. Kort-Kamp, B. Amorim, G. Bastos, F. A. Pinheiro, F. S. S. Rosa, N. M. R. Peres, and C. Farina, Phys. Rev. B {\bf 92}, 205415 (2015).
\bibitem{Sounas_2012}  D. L. Sounas and C. Caloz, IEEE Trans. Microw. Theory Tech. {\bf 60}, 901 (2012).

%Condutividade Grafeno
\bibitem{Gusynin1}  V. P. Gusynin, and S. G. Sharapov, Phys. Rev. B {\bf 73} 245511 (2006).
\bibitem{Gusynin2} V. P. Gusynin, S. G. Sharapov, and P. Cabotte, J. Phys.: Cond. Mat. {\bf 19}, 026022 (2007);

\end{thebibliography}
\end{document}